\documentclass[journal, english, 10pt, twocolumn, twoside]{IEEEtran}

\usepackage[numbers,sort&compress]{natbib}

\ifCLASSINFOpdf
  \usepackage[pdftex]{graphicx}
  \usepackage{amsmath}
  \usepackage{import}
  \usepackage{multirow}
  \usepackage{amssymb}
  \usepackage{hyperref}

\hypersetup{hidelinks,
	colorlinks=true,
	allcolors=black,
	pdfstartview=Fit,
	breaklinks=true}

\else

\fi

\usepackage{amsmath}

\usepackage{algorithmic}

\usepackage{array}

\usepackage{url}
\usepackage{color}

\newcommand{\eg}{\emph{e.g.},~}
\newcommand{\etal}{\emph{et al}.~}

\begin{document}

\title{Physics-Informed Ensemble Representation for Light-Field Image Super-Resolution}

\author{
Manchang~Jin\IEEEauthorrefmark{2},~Gaosheng~Liu\IEEEauthorrefmark{2},~Kunshu~Hu,~Xin~Luo,~Kun~Li,~\IEEEmembership{Member, IEEE},~Jingyu~Yang,~\IEEEmembership{Senior Member, IEEE}

\thanks{
This work was supported in part by the National Natural Science Foundation of China under Grant 62231018 and Grant 62171317. (\emph{Corresponding author: Jingyu Yang})

M. Jin, G. Liu, K. Hu, X. Luo, and J. Yang  are with the School of Electrical and Information Engineering, Tianjin University, Tianjin 300072, China.

Kun Li is with the College of Intelligence and Computing, Tianjin
University, Tianjin 300350, China.

\IEEEauthorrefmark{2}These authors contributed equally to this work.}}

\def\etal{\emph{et al}.~}

\maketitle

\begin{abstract}
Recent learning-based approaches have achieved significant progress in light field (LF) image super-resolution (SR) by exploring convolution-based or transformer-based network structures. However, LF imaging has many intrinsic physical priors that have not been fully exploited. In this paper, we analyze the coordinate transformation of the LF imaging process to reveal the geometric relationship in the LF images. Based on such geometric priors, we introduce a new LF subspace of virtual-slit images (VSI) that provide sub-pixel information complementary to sub-aperture images.
To leverage the abundant correlation across the four-dimensional data with manageable complexity, we propose learning ensemble representation of all $C_4^2$ LF subspaces for more effective feature extraction. To super-resolve image structures from undersampled LF data, we propose a geometry-aware decoder, named \emph{EPIXformer}, which constrains the transformer’s operational searching regions with a LF physical prior. Experimental results on both spatial and angular SR tasks demonstrate that the proposed method outperforms other state-of-the-art schemes, especially in handling various disparities. Our codes are available at 
\href{https://github.com/kimchange/PILF}{\textcolor[rgb]{1.0,0.10,0.80}{https://github.com/kimchange/PILF}.}
\end{abstract}

\begin{IEEEkeywords}
Light field image processing, feature representation, super resolution, transformer, physical modelling.
\end{IEEEkeywords}

\IEEEpeerreviewmaketitle

\section{Introduction}\label{intro}

\IEEEPARstart{L}{ight} field (LF) imaging can capture both the 
intensities and directions of light rays that pass through a scene.
Compared with conventional 2D imaging, it provides rich geometric information for scene understanding and consequently enables a variety of applications such as  semantic segmentation~\cite{SHENG2022CSVT}, depth estimation \cite{PAN2017TMM}, 3D rendering \cite{NG2012TMM}, object detection \cite{CHEN2023TMM} and so on.However, LF imaging suffers from an intrinsic tradeoff between spatial and angular resolutions when encoding the 4D structure into a 2D sensor due to the limited resolution of imaging sensors.
To handle the resolution dilemma, developing computational algorithms to enhance the spatial and angular resolutions is drawing increasing attentions. 
In the literature, two principal tasks are introduced to address the problem: spatial super-resolution (LFSSR) and angular super-resolution (LFASR). One key in common is incorporating the 4D LF structural prior into the method design. 

\begin{figure}[t]
    \centering
    \includegraphics{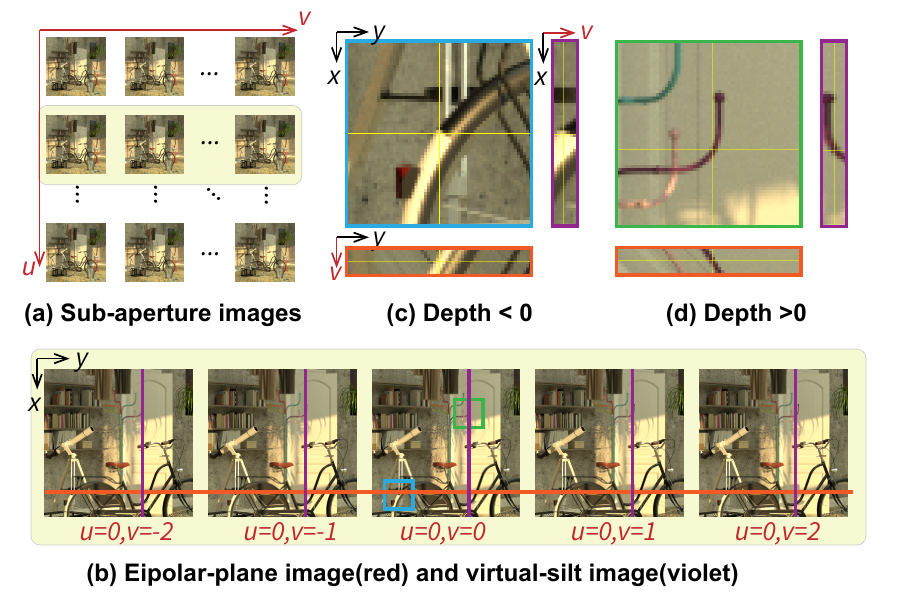}
    \caption{An illustration of VSI formation: (a) SAIs, each image represents observation from a different angle. (b) One row of SAIs, the red line selects one horizontal EPI and the violet line selects one vertical VSI, note that this selection is just for illustration and not unique. Then we zoom in the blue region (c) where depth $<$ 0 and the green region (d) where depth $>$ 0. Note that all sub-images in (c) and (d) are shown in the \textbf{\textit{same}} scale factor. We can find that VSI can produce cognizable image with a different sampling factor, which can be interpreted as observing the scene from a virtual slit thus may contain sub-pixel information compared to SAI.
    }
    \label{fig:C42 representation}
\end{figure}

The traditional optimization-based methods designed for LF super-resolution exploited non-local information to recover the LF, such as LFBM5D \cite{alain2018light}, whose PSNR improved about 2dB than bicubic interpolation. Benefiting from the success of convolutional neural networks (CNNs), the learning-based methods for LFSR have made significant progress in reconstructing the high-resolution (HR) LFs from low-resolution (LR) observations. Learning-based method RCAN outperforms LFBM5D without considering the light-field structural prior, which implies great potential for learning-based LF super-resolution approaches. Among them, various modules have been introduced to utilize domain-specific 2D convolutions to extract the structure prior from 4D LF data. 
For example, spatial-angular alternating convolution~\cite{Yoon}, spatial-angular versatile convolution~\cite{cheng2022spatial}, and disentangling mechanism~\cite{Wang_2023}.
In summary, these methods consider performing convolutions on the subset of three well-known 2D representations of LF images, \textit{i.e.}, sub-aperture images (SAIs), macro-pixel images (MacPIs), and epipolar-plane images (EPIs), which are experimentally effective in terms of their reported results.
However, based on the analysis of the coordinate transformation of the LF imaging process,  
we found that LF imaging is essentially a 4D imaging with significant 2D constraints (see Section~\ref{sec:motivation}), and there still exists other important priors that can directly constrain the network design and regularize the training. And based on this observation, we re-organize LF image into a new cognizable 2D representation, 
termed as virtual-silt images (VSIs), as illustrated in Fig.~\ref{fig:C42 representation} (c) and (d).
The VSI can be regarded as observing the scene from a single slit, which provides a new perspective to represent an LF and can contain rich sub-pixel information at that specific line position. More analyses of VSI can be found in section \ref{C42 representations for 4D LF}.

Based on these observations, we leverage the SAI, MacPI, EPI, and VSI representations and consequently propose a $C_4^2$ feature extractor ($C_4^2$FE) to fully exploit the structure prior of LF images.
The $C_4^2$FE also unifies previous modules for LF feature extraction by complementing the VSI feature extraction into our unified design to make it a complete set. And since $C_4^2$ is designed as a generic extractor for all tasks that fulfill ``4D data with 2D priors", we further designed a LF specialized, transformer-based architecture,  EPIXformer as a context-aware decoder, which is constrained according to the LF imaging character to further enhance the capability to handle large disparities.

We apply the $C_4^2$FE and EPIXformer on both LFSSR and LFASR tasks, 2 well-known LF challenges to prove the effectiveness of our design. Experimental results show that our methods achieved better visual and quantitative results than previous studies.

Our main contributions are as follows: 
\begin{itemize} 
    
    \item We analyzed the coordinate transformation 
 to reveal geometric correlation in LF imaging, and introduce a new LF subspace of virtual-slit images (VSI) that provide sub-pixel information complementary to sub-aperture images. Based on this analysis, we propose learning ensemble representations derived from $C_4^2$ LF subspaces to exploit the abundant correlation across the 4D data.
    \item We propose a geometry-aware decoder, named \emph{EPIXformer}, that constrains the transformer’s operational searching regions by the geometric trajectories of the scene’s LF projection. This facilitates robust recovery of image structures from undersampled LF data with various disparities.

\item Compared with state-of-the-art methods, extensive experimental results demonstrate that our method outperforms other methods in terms of quantitative and visual performance on public datasets for both LFSSR and LFASR tasks, verifying the effectiveness of the proposed physics-informed learning framework for LF images.

\end{itemize}

\section{Related Work}
This section briefly reviews the related studies on LF spatial and angular SR.
\subsection{LF Spatial SR}
The LFSSR aims at generating high-resolution (HR) LF images from their low-resolution (LR) counterparts. Previous methods for LFSSR can be broadly categorized into optimization-based and learning-based approaches. 

Optimization-based approaches typically leverage disparity information as priors to aid the SR process. For example, Bishop \etal\cite{Bishop} utilized a variational Bayesian framework to reconstruct the scene disparity for LFSSR. Wanner \etal\cite{Wanner} initially estimated disparity maps using EPIs, then utilized them to generate HR images. Mitra \etal\cite{Mitra} proposed to process LF patches using a Gaussian mixture model based on disparity values. Rossi \etal\cite{Rossi} introduced a graph-based regularizer that enforces the LF geometric structure, enabling the utilization of complementary information across different views for LFSSR. However, these methods heavily rely on accurate depth or disparity information, which can easily lead to unsatisfactory results due to inaccurate estimations.

In recent years, learning-based methods have been introduced for LFSSR \cite{Yoon, Yuan, Yeung, wang2020spatialangular, Wang_2023, Jin, Cheng2020CSVT, MO2022CSVT, ZHOU2023CSVT} driven by the CNNs. Yoon \etal\cite{Yoon} proposed the pioneer method to simultaneously achieve angular and spatial super-resolution. Yuan \etal\cite{Yuan} first employed a single-image SR method on each SAI and then developed a deep EPI-based neural network to further reconstruct the geometric structure of LF. Cheng \etal\cite{Cheng2020CSVT} utilized internal and external view similarities using a CNN, Mo \etal\cite{MO2022CSVT} further used an attention strategy to enhance the performance. Yeung \etal\cite{Yeung} utilized spatial-angular separable convolutions instead of 4D convolution to save memory and computational costs, which can efficiently extract spatial and angular joint features. Zhou \etal\cite{ZHOU2023CSVT} combined the depth defocus information into a CNN-based network. Wang \etal\cite{wang2020spatialangular} proposed LF-InterNet to progressively interact spatial and angular features. Later, they further designed disentangling mechanism \cite{Wang_2023} to extract four domain-specific features for LFSSR. Jin \etal\cite{Jin} developed LF-ATO, which first adopted an all-to-one strategy for SR and then utilized structure regularization to refine the result. Liang \etal\cite{LFT,EPIT} used transformer on spatial, angular and EPI domain to deal with LF iamges with large disparity. These methods provide important examples to incorporate observations into network design. Since structure determines function, all observations can be reflected on the physical imaging process, thus the performance can be further improved by incorporating physical priors.

\subsection{LF Angular SR}
The LFASR aims at synthesizing novel SAIs from given views, enriching the angular information of LF images.
Previous works for LFASR could generally be categorized into depth-based and non-depth-based methods.

Depth-based methods synthesize novel views with the help of disparity estimation. Pearson \etal\cite{Pearson} used plenoptic sampling theory as guidance to obtain layer-based geometric information of objects, which helps to synthesize novel views at arbitrary angular positions. 
Wanner \etal\cite{Wanner} used EPI analysis with convex optimization techniques to estimate subpixel-level disparity maps, then calculate the warping maps required for novel view synthesis. 
Kalantari \etal\cite{Kalantari} first used CNNs to estimate disparity maps and synthesize new SAIs. 
Jin \etal\cite{Jin2020LearningLF} used a disparity estimator with a large receptive field to estimate depth maps with a blending strategy to refine the results.
Jin \etal\cite{Jin_Deep2022} also proposed to construct plane-sweep volumes (PSVs) to achieve more accurate estimation. 
Ko \etal\cite{Ko} proposed an adaptive feature remixing method to realize angular SR. The depth-based methods require extra costs to estimate the disparity, and the ASR accuracy often relies much on disparity accuracy.

For non-depth-based methods, Vagharshakyan \etal\cite{Vagharshakyan} used shearlet transform to provide a sparse representation of the EPIs for LF reconstruction. 
Wu \etal\cite{Wu} trained CNNs to select sheared EPI patches and fused sheared EPIs to reconstruct high-angular-resolution LF images. 
Yeung \etal\cite{Yeung_2018_ECCV} reconstructed a densely-sampled LF from a sparsely-sampled LF by alternating convolutions on spatial and angular dimensions. 
Wu \etal\cite{WuG} designed a deep learning pipeline to handle the aliasing, similar to the conventional Fourier reconstruction filter. 
Later they proposed a spatial-angular attention network, SAA-Net \cite{WuG2} to cope with non-Lambertian surface and large disparity.
Sheng \etal\cite{SHENG2023CSVT} used a recurrence-based method to recover the angular resolution.
Wang \etal\cite{Wang_2023} proposed DistgASR, using SAIs, MacPIs and EPIs subspace to disentangle LF features.
Liu \etal\cite{Liu} achieved ASR by using MacPIs upsampling, which provides an efficient structure to recover the angular resolution. These non-disparity-based methods struggle to handle LFs with large disparity. 
\section{Motivation}
\label{sec:motivation}
\subsection{4D LF Optical Imaging}
The forward LF imaging model can be formulated as the following linear model.
\begin{equation}
\label{eq:genericLF}
\begin{aligned}
I(x, y, u,v) = \int S(x,y,z) \ast h(u,v,x,y,z) \mathrm{d}z,
\end{aligned}
\end{equation}
where $\ast$ is the 2D convolution operated on dimension $(x,y)$, it means the 4D LF image $I$ is the convolution of the 3D scene $S$ and the 5D point spread function (PSF) $h$, and  coordinate indices include: $u$ and $v$ for two angular dimensions, $x$ and $y$ for  two lateral dimensions, and $z$ for the axial dimension.

\begin{figure}
    \centering
    \includegraphics[width=1.0\linewidth]{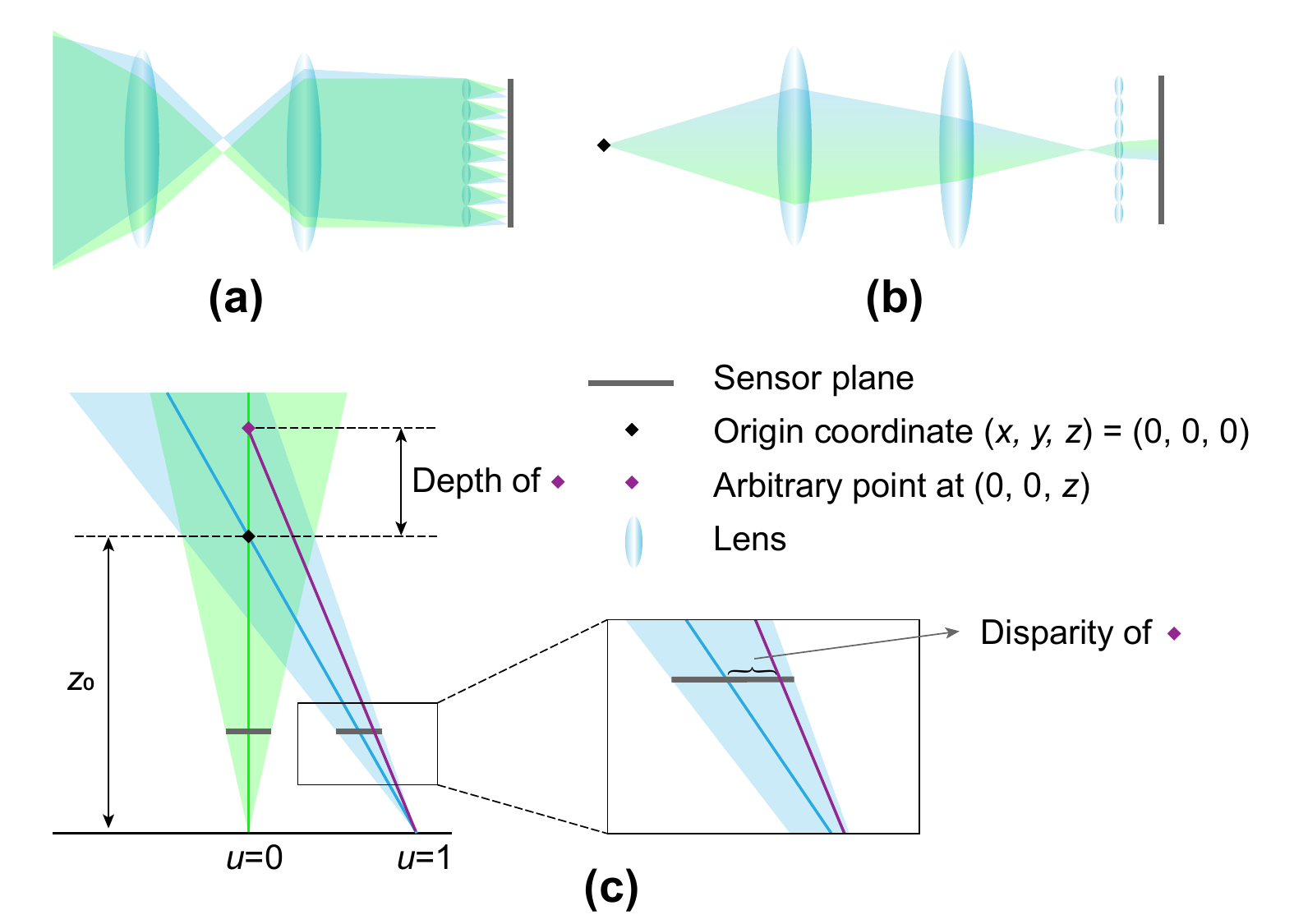}
    \caption{The analysis of LF imaging, the blue or green color represents the light path of 2 different angles. (a) The light beam path of SAI. (b) The light beam path of MacPI. (c) The coordinate mapping from real-world space to image space. }
    \label{optical_path}
\end{figure}

Through the imaging formation (\ref{eq:genericLF}), the characteristics of LF images not only depend on the latent scene content but are also highly modulated by the system's PSF. We first exploit structures of LF PSF to inspire the design of efficient feature extraction for 4D LF images. And geometric-optics dominate the scenes where pixel size is far greater than the wavelength of visible light, 
thus like \cite{ng2005fourier}, we simplify the LF PSF into a delta function to focus on the sampling coordinates from the real-world to the imaging plane for each angle.
Assuming Lambertian surface and leaving out  out-of-focus blurring, and based on the geometry analysis of Fig~\ref{optical_path}, the LF PSF can be approximated as
  \begin{equation}
  \label{PSF}
  \hat{h} \left(u,v,x,y,z\right)=
  s^2(z)\delta\left(s(z)x-uz/z_0, s(z)y-vz/z_0)\right),
  \end{equation}
where $\delta$ is the delta function. Define the origin of $z$ axis by its intersection with the focused plane. Denote by $z_{0}$ the focus distance between the camera and the focused plane, at which the disparity is 0. Then, $s(z) = 1 + z/z_0$ is the scaling factor depending on the object position relative to the focused plane.
Angular coordinates $(u,v)$ control the disparity based on axial position $z$. The center position is defined as the origin of the 4D LF coordinates $u,v,x,y$. The world coordinate of the camera position is $(0,0,-z_0)$, and $z \in [-z_0, +\infty)$.
As a result, the LF imaging can be simplified as 

\begin{equation}
\label{eq:simpleLF}
\begin{aligned}
\hat{I}(x,y,u,v)&= \int S(x,y,z) \ast \hat{h}(u,v,x,y,z) \mathrm{d}z \\
&=\int S\left(s(z)x-uz/z_0, s(z)y-vz/z_0, z\right) \mathrm{d}z,
\end{aligned}
\end{equation}

\subsection{\texorpdfstring{$C_4^2$ 2D Subspaces of 4D LF images}{C42 2D Subspaces of 4D LF images}}\label{C42 representations for 4D LF}
From Eq.~(\ref{eq:simpleLF}), the coordinate transformation and the projection along the $z$ dimension multiplex 3D information of scene $S$ into the 4D LF space, which indicates that the four dimensions of the LF space are highly correlated. Since $u,v,x,y$ only occupy the first and second dimensions of scene $S$, we think that 2D constraints are dominant. On the other hand,  directly exploiting the four-dimensional (4D) correlation would impose intensive computational and memory demands. Instead, following the divide-and-conquer philosophy, previous approaches turns to exploit correlation of 2D subspaces~\cite{Yoon}, \eg
the $x$-$y$ SAI domain \cite{Yoon}, $u$-$v$ MacPI domain \cite{wang2020spatialangular}, and $u$-$x$ and $v$-$y$ EPI domains \cite{Wang_2023}. Note that, among the $C_4^2=6$ subspaces, 
previous works seldom explored the $u$-$y$ and $v$-$x$ domains, which might miss prominent LF priors. 

Fig. \ref{fig:C42 representation} visualizes three 2D subspace images of an LF image: SAIs, EPIs, and VSIs on the $x$-$y$,  $v$-$y$, $v$-$x$ domains, respectively. The characteristics of SAIs and EPIs have been recognized early on: SAIs are images observed at different angles, while EPIs are linear patterns with depth-related slopes. Note that  $u$-$y$ ($v$-$x$) images are stacks of co-located SAI rows (columns) along the orthogonal direction in the angular domain, which are essentially images captured by tilting a virtual line camera around a virtual slit at a different depth.
Therefore, we named them as \emph{virtual slit images} (VSI). 
Hence, intuitively, a VSI contains similar image structures to the associated local region of the SAI. 
According to the coordinate transformation in the imaging model (\ref{eq:simpleLF}), the pixel sampling stride along the horizontal lateral dimension of the $x$-$y$ SAI is $s(z)$,
while that of the $x$-$v$ VSI is $-z/z_0$. 
The $x$-$v$ ($u$-$y$) VSIs have different sampling rates from the $x$-$y$ SAIs for the horizontal (vertical) lateral dimension. Therefore, with different sampling rates, VSIs can provide sub-pixel information complementary to SAIs. 
If an object is at depth $z_\textrm{vsi}$, then its VSI sampling rate is the same to SAI sampling rate at depth $z_\textrm{sai}$, to see the relationship, 
let $s(z_\textrm{sai}) = -z_\textrm{vsi}/z_0$, 
and we have 
\begin{equation}
\label{equ:SamRate}
z_\textrm{sai} = -z_\textrm{vsi} -z_0.
\end{equation}
Eq.~(\ref{equ:SamRate}) indicates that VSIs are equivalent to monocular imaging by moving the object of depth $z_\textrm{vsi}$ to a new depth $z_\textrm{sai}$. For example, when observing an object at $z=0$, VSI is equivalent to monocular imaging by moving the object to $z=-z_0$, at which it would ideally have an infinite sampling rate. 

The distance between the camera at $-z_0$ and the object at $z_\textrm{sai}$ is $z_\textrm{sai} - (-z_0) = -z_\textrm{vsi}$, which means that for an object at $z_\textrm{vsi}$, we can also treat its VSI to be the reversed observation of a virtual silt at $z=0$. We can also see it from Fig~\ref{optical_path} (c), a green line and the blue line will be two adjacent lines on DPI and they converge on the same line at depth $z=0$, which is the position of virtual silt described above. 

Based on the above analysis, with depth-dependent sampling rates, the proposed VSI representation contains much sub-pixel information relative to the SAI subspace, especially when the disparity is relatively small, so leveraging the VSI information would benefit the processing of LF images. For the 4D LF image space, there are six ($C_4^2$) subspaces in total: two for VSI, two for EPI, one for SAI and the rest one for MacPI, respectively. In this work, we propose learning ensemble representation of all $C_4^2$ LF subspaces, termed $C_4^2$ LF representation for short hereafter, to extract LF features more thoroughly.

\section{\texorpdfstring{$C_4^2$ LF Feature Extractor and EPIXformer}{C42 LF Feature Extractor and EPIXformer}}

Based on the analysis in Section III, we designed two main components, namely $C_4^2$ feature extractor and EPIXformer, for LF feature extraction and decoding.
We apply the proposed modules for both LFSSR and ASR tasks.
In what follows, we describe the design and structure of the two modules.
\subsection{\texorpdfstring{$C_4^2$ LF Feature Extractor}{C42 LF Feature Extractor}}

Previous studies have demonstrated the power of CNNs in the representation and processing of 4D LF images, and 2D convolutions with elaborated network structures can outperform the straightforward 4D convolutions~\cite{Wang_2023}. Based on the above analysis, we design a backbone network for efficient feature extraction to learn the ensemble representation of $C_4^2$ LF subspaces, including $x$-$y$ SAI (spatial), $u$-$v$ MacPI (angular), $x$-$u$ EPI (vertical), $v$-$y$ EPI (horizontal), $x$-$v$ VSI (vertical), $y$-$u$ VSI (horizontal).

\begin{figure}[t]
    \centering
    \includegraphics[width=0.9\linewidth]{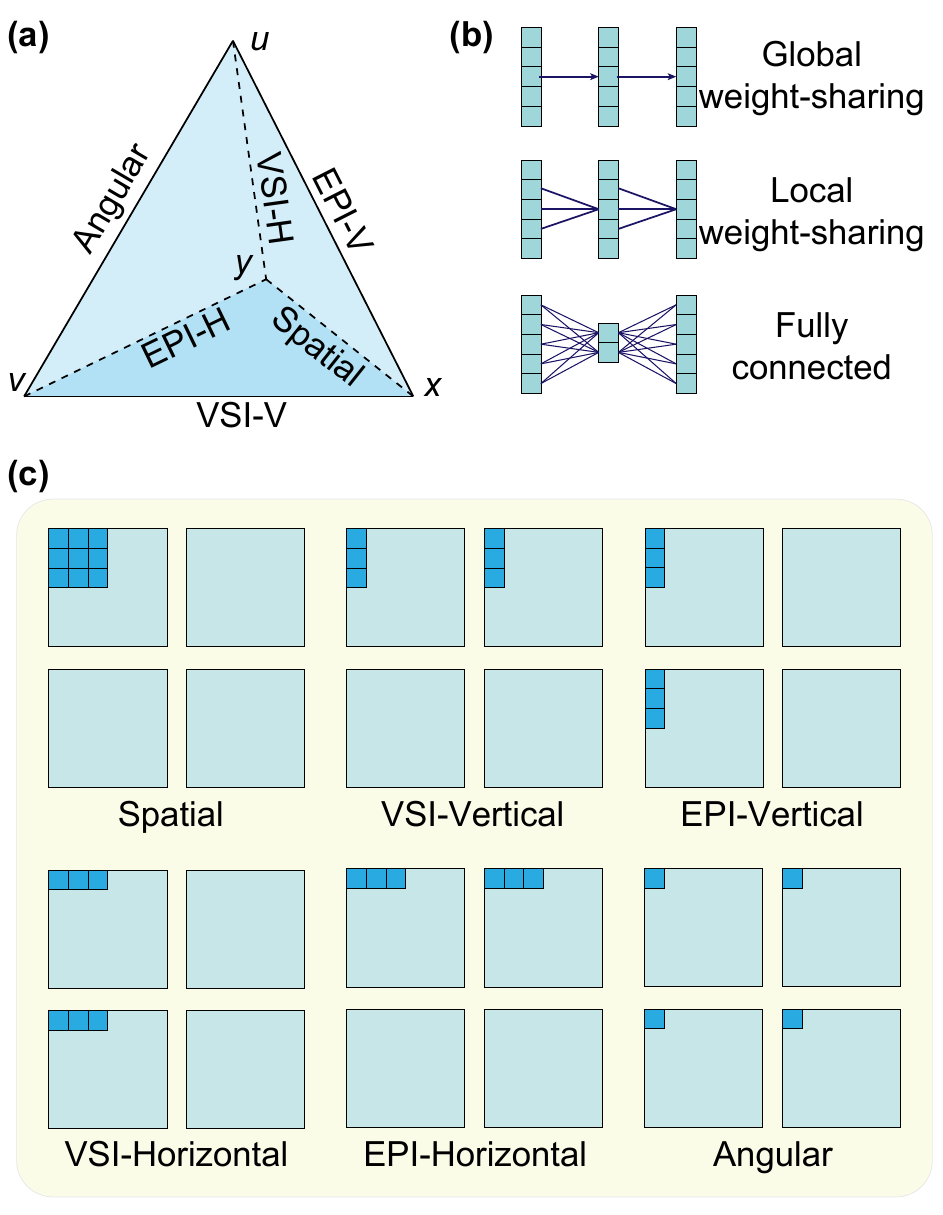}
    \caption{An illustration of the proposed $C_4^2$ feature extractor: (a)  $C_4^2$  feature extractor chooses 2 dimensions at a time from 4 LF dimensions $u,v,h,w$. (b) Global weight-sharing is applied on unchoosed dimensions. For chosen dimensions, depending on the dimension's element number, local weight-sharing is applied on chosen spatial dimensions $h,w$ , and chosen angular dimensions $u,v$ are fully connected. (c) The resulting $C_4^2$ feature extractor appears on LF SAIs in a convolutional layer form, for illustration.  }
    \label{C42 LF extractor}
\end{figure}

Different from disentangling block~\cite{wang2022learning}, which utilizes domain-specific convolutions to first downsample the MacPI and EPI features into subspace and then upsample them into the original size, we adopt fully convolutional layers to equivalently realize these modules.
Fig.~\ref{C42 LF extractor} (c) illustrates how the convolution kernel operates on the input to extract features from the corresponding subspace. Based on these subspace-specific convolutions, as shown in Fig.~\ref{network-overview} (b), we design a feature extraction module, named \emph{$C_4^2$ Conv block}, to learn the ensemble of $C_4^2$ subspace representation.  In this module, the feature maps generated by subspace-specific convolutions are activated by two LeakyReLUs separated by a 1$\times$1 convolution, and  followed by a channel to angle (C2A) rearrangement, if necessary, so that the output features of each extractor have the same shape. Then, the features from the six branches are concatenated and fused by the cascading of $1\times1$ convolution, a Leaky ReLU activation, and a $3\times3$ convolution.  In the $C_4^2$ Conv block, residual learning is applied by using a local skip connection.

\subsection{EPIXformer as a Geometry-Aware Decoder}
\begin{figure*}[t]
    \centering
    \includegraphics[width=0.9\linewidth]{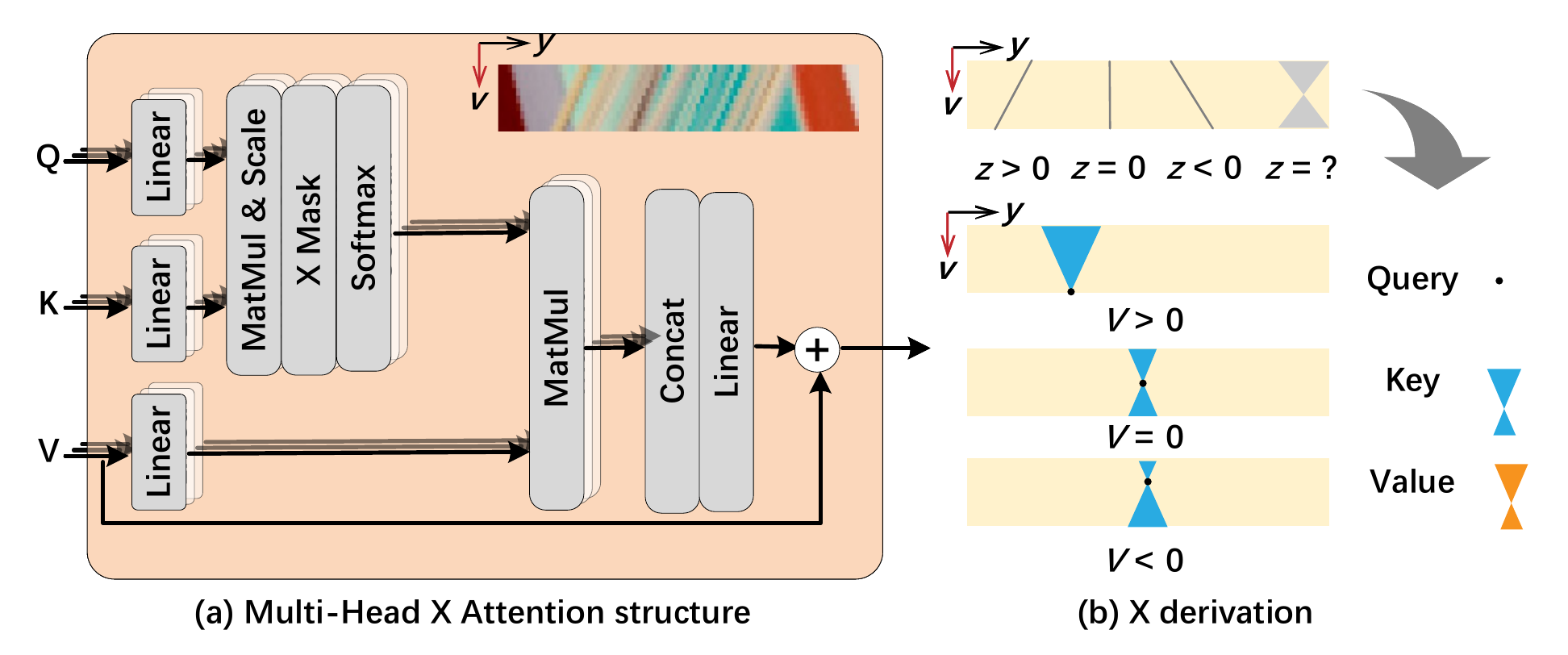}
    \caption{Structure for proposed MHXA: (a) An EPI from a LF image (b)  Image function without occlusions using our equation, showing that when depth is unknown in advance, all possible lines on EPI make an 'X' shape, which determines the designed search window of Query to Key. We listed several 'X' shape of Query at different locations.}
    \label{EPIXformer}
\end{figure*}

Some previous LFSSR methods achieved promising performance by using transformers
on spatial, angular, or EPIs~\cite{LFT,EPIT} domains to exploit non-local information. As a departure, in this work, we design the transformer with the geometric structures of LF imaging. The basic observation is that, given an undersampling LF configuration, a 3D point missed in some angles might be sampled in others due to the disparities among SAIs, and pixels to be super-resolved can be predicted by the sampled counterparts from neighbouring angles with geometric priors. 

As analyzed in Sec. \ref{sec:motivation}, LF images have multiple domain-specific representations, and we should select the one where LF geometric characteristics can be well captured by the transformer framework. As indicated in Eq. (\ref{eq:simpleLF}), of the six subspaces, the $u$-$x$ or $v$,$y$ dimensions of the EPI subspaces are multiplexed in the same dimension of the latent scene. Thus, there are many pixels that capture the same physical position in the EPIs. Therefore, we operate transformers on the EPI representation similar to \cite{EPIT}. Specifically, consider an $x$-$u$ EPI at particular $\bar{v}$ and $\bar{y}$, LF pixels observing an unoccluded 3d point $(x',y',z')$ of the latent scene $S$ are determined 
\begin{equation}
    u = \left( (z'+z_0) x - z_0  x' \right)/z'.  
    \label{equation:EPIXformer}
\end{equation}
Similarly, the observed pixels in an $v$-$y$ EPI  at particular $\bar{x}$ and $\bar{u}$ are given by 
$v = \left( (z'+z_0) y - z_0  y' \right)/z'$.
As a result, image features in EPIs are distributed along lines with slope $d=(z'+z_0)/z'$ as disparity between associated angles. Such a strong geometric prior provides useful clues for recovering missing information, particularly in the presence of severe degradation.

Specifically, as shown in Fig~\ref{EPIXformer}(b),
for an unknown depth $z$, all possible pixels capturing the same 3D position form an ``X'' shape. To leverage these geometric structures, 
we propose a transformer structure, named EPIXformer, by constraining its operational searching region to the ``X'-shape areas. The structure of EPIXformer is shown in Fig.~\ref{EPIXformer}(a). Taking the horizontal EPIXformer as an example, it takes pixel-level features as token $T\in\mathbb{R}^{UH\times VW\times C}$ to generate query $\mathcal{Q}$, key $\mathcal{K}$, and value $\mathcal{V}$ as follows.
\begin{equation}
    \begin{aligned}
        \mathcal{V} &= T\\
        \mathcal{Q} = \mathcal{K} &= \mathrm{LayerNorm}(T).
    \end{aligned}
\end{equation}
Denote by $N_H$ the number of heads. We split $C$ channels into $N_H$ groups to perform multi-head self-attentions~\cite{vaswani2017attention}.
For each head, EPIXformer constrains the search window of $\mathcal{Q}$ based on Eq.~(\ref{equation:EPIXformer}). 
Since the slopes of linear structures in EPIs are unknown in real scenarios, we propose using an X-shape search window that adapts to possible depth in a wide range.

Specifically, denote by $\mathrm{H}$ the output of a particular head, and by  $\mathrm{W}_q$, $\mathrm{W}_k$, and $\mathrm{W}_v$  the learnable matrices of multiplication projection weights for query $\mathcal{Q}$, key $\mathcal{K}$, and value $\mathcal{V}$, respectively. The transformer of the  $h^\textrm{th}$ head is described as
\begin{equation}
\begin{aligned}
    \mathrm{H} &= \mathrm{X} \mathcal{V} \mathrm{W}_{v}, \\
    \mathrm{X} &= \mathrm{Softmax}\left(\frac{\mathcal{Q} \mathrm{W}_{q} (\mathcal{K} \mathrm{W}_{k})^\top }{\sqrt{C/N_H}} + \mathrm{M}\right),
\end{aligned}
\end{equation}
where $\mathrm{M}$ is the mask to constrain the operational regions defined as
\begin{equation}
    \mathrm{M}(v_q, y_q, v_k, y_k) = 
\begin{cases}
0,  & \text{if } \left | \frac{ y_k -y_q }{v_k-v_q}  \right | \le  d_\textrm{max},  \\[2ex]
-\infty, & \text{otherwise}.
\end{cases}
\label{eq:mask}
\end{equation}
In Eq. (\ref{eq:mask}), $y_k$ and $v_k$ ( $y_q$ and $v_q$) are indices for the key (query) matrix, and $d_\text{max}$ is maximum disparity depending LF task configuration (detailed in Sec. \ref{subsec:Network architecture}).

Denote by $\mathcal{H} = \mathrm{\left [ H_1, H_2,...,H_{N_h} \right ]}$ the concatenation of all heads. The output feature of all the heads is $\mathcal{H}\mathrm{W_O}$, where $\mathrm{W_O}$ is the learnable multiplication projection matrix. 
The output of the MHXA block is obtained by using a residual update strategy $T := \mathcal{H}\mathrm{W_O} + T$ as shown in Fig~\ref{EPIXformer}(a). 

\begin{figure*}[t]
    \centering
    \includegraphics[width=1.0\linewidth]{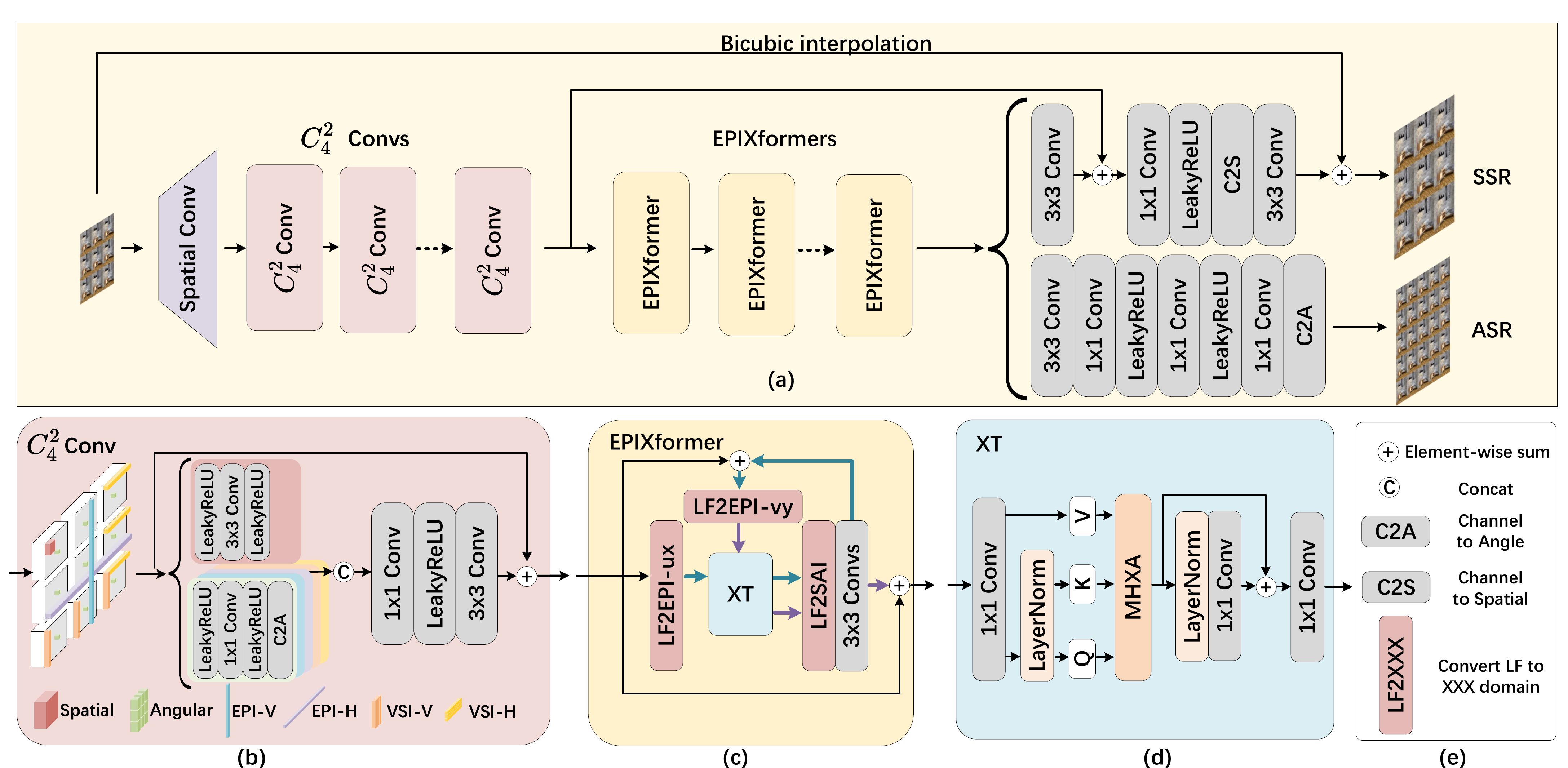}
    \caption{Overview of proposed spatial and angular super-resolution networks.}
    \label{network-overview}
\end{figure*}

With the proposed MHXA block, the constructed transformer (XT) is shown in Fig~\ref{network-overview}(d). As shown in Fig.~\ref{network-overview}(c), the overall transformer structure,  named \emph{EPIXformer}, operates XT twice successively on horizontal and vertical EPI subspaces to exploit the geometric correlation. The feature is finally converted to the SAI domain to have the same shape as its input.

\begin{figure*}[htbp]
    \centering
    \includegraphics[width=1.0\linewidth]{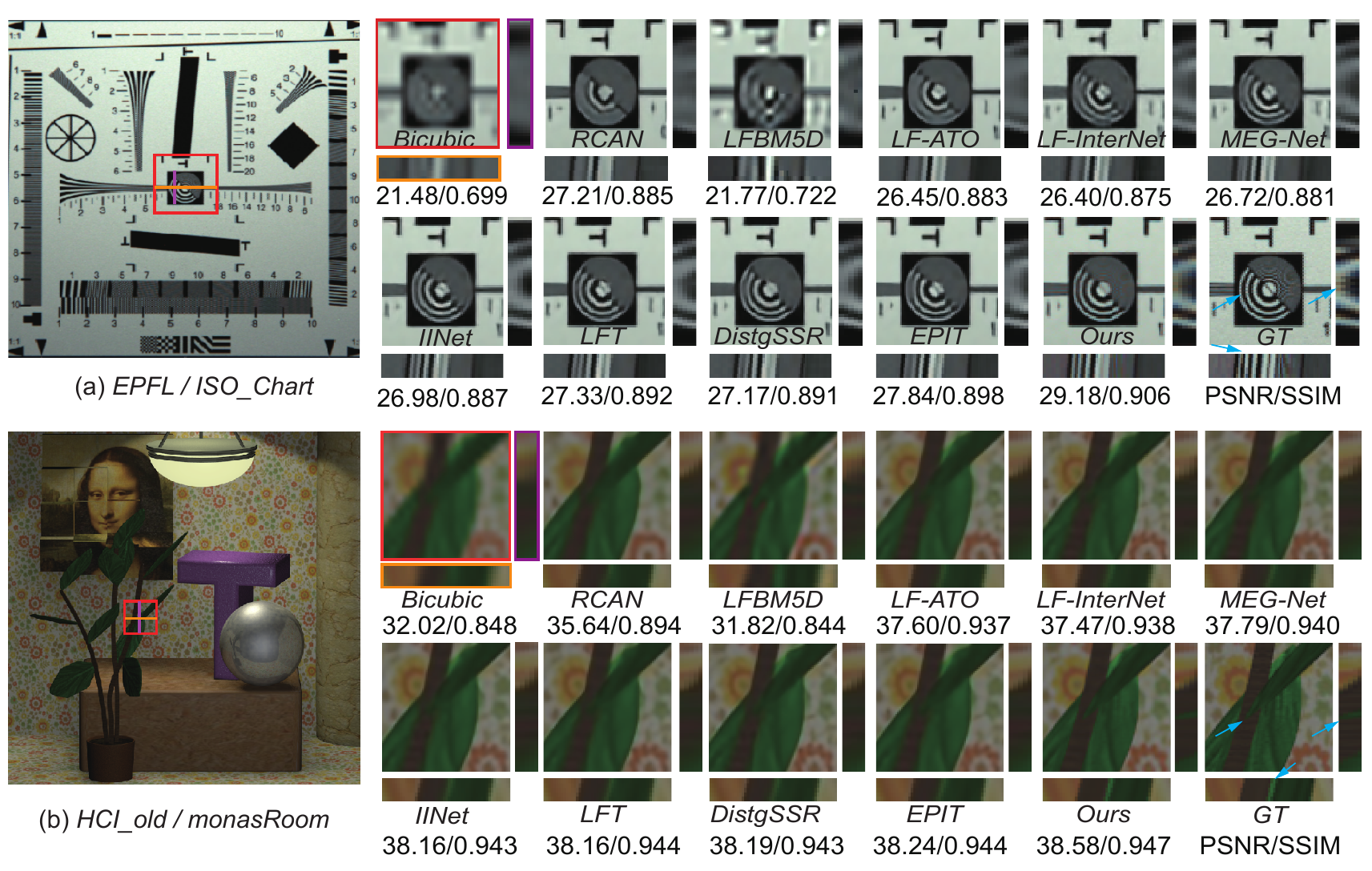}
    \caption{Visual performance on LF spatial super resolution $\times 4$ : (a) The center view (red), the vertical VSI (violet) and the horizontal  EPI (orange) of all methods. (b) The center view (red), the vertical VSI (violet) and the horizontal  EPI (orange) of all methods. 
    }
    \label{SSRx4figure}
\end{figure*}

\begin{figure*}[htbp]
    \centering
    \includegraphics[width=1.0\linewidth]{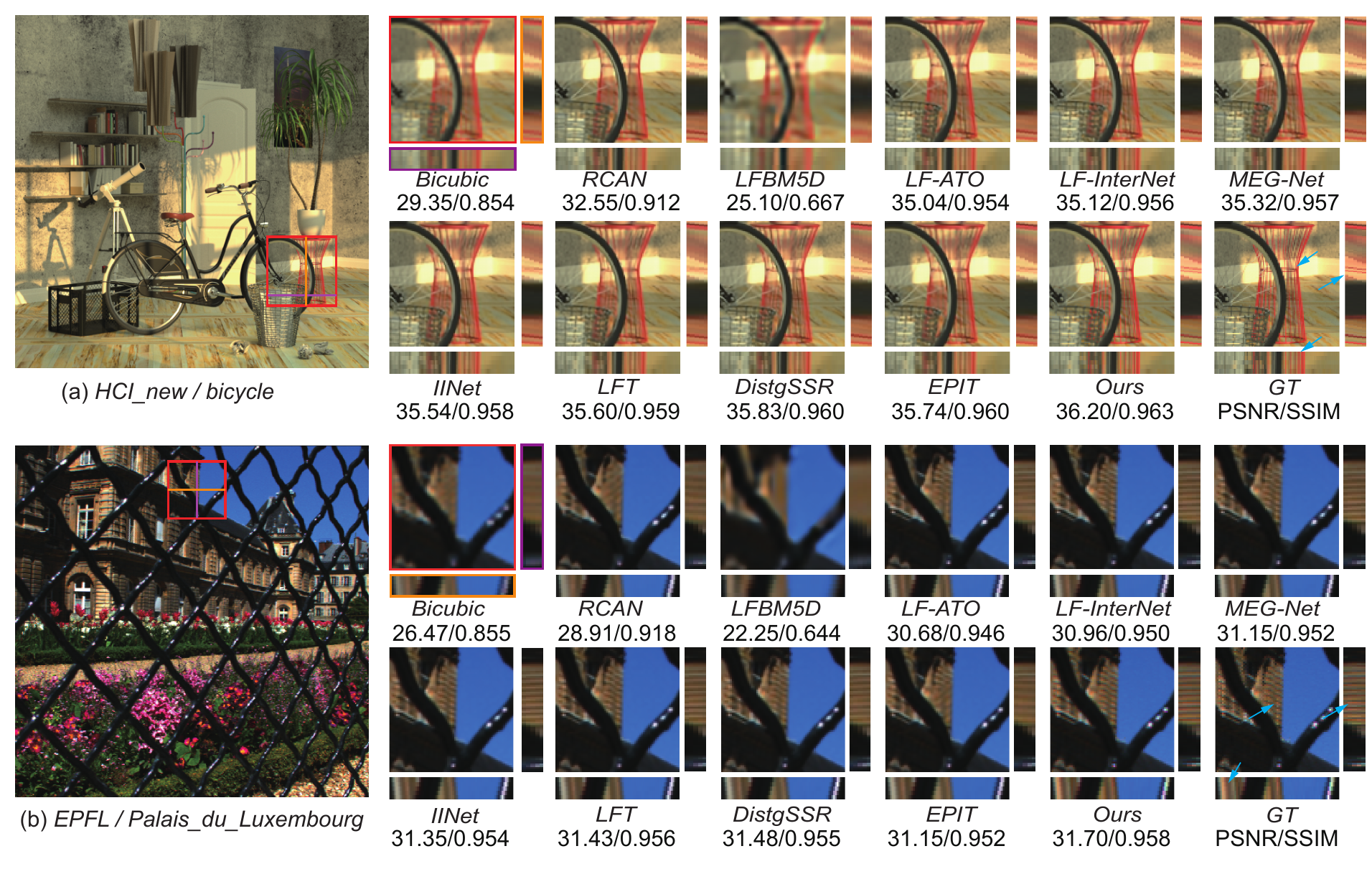}
    \caption{Visual performance on LF spatial super resolution $\times 2$, the order of these comparing images is the same to $\times 4$ : 
    (a) The center view (red), the horizontal VSI (violet) and the vertical EPI (orange) of all methods.  
    (b) The center view (red), the vertical VSI (violet) and the horizontal  EPI (orange) of all methods.}
    \label{SSRx2figure}
\end{figure*}

\begin{figure*}[htbp]
    \centering
    \includegraphics[width=1.0\linewidth]{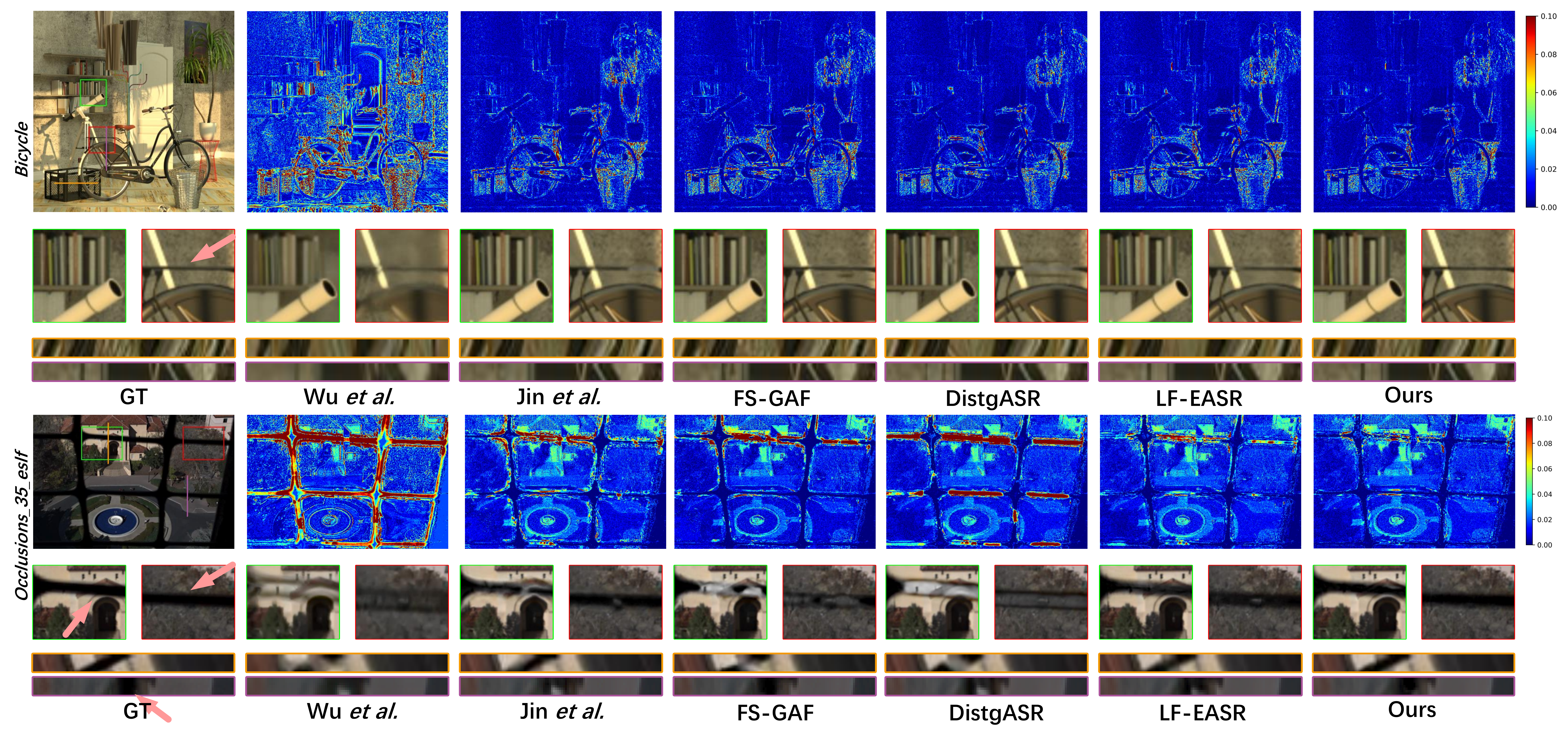}
    \caption{Visual comparison on real-world scenes for $2\times2\rightarrow7\times7$ task. For comparison, we provide the error maps, enlarged patches highlighted in red and green boxes, and VSI (cutting along the violet line ) and EPI (cutting along the orange line) of all methods. }
    \label{VisualASR}
\end{figure*}

\begin{table*}[htbp]
\caption{Quantitative comparison of different SR methods in terms of PSNR/SSIM. The best results are shown in bold.}
\label{SSR-psnr-ssim}
\renewcommand\arraystretch{1.60}
\resizebox{\linewidth}{!}{
\begin{tabular}{|c|ccccc|ccccc|}
\hline
\multirow{2}{*}{Methods} & \multicolumn{5}{c|}{$\times$2}                                                                                                                         & \multicolumn{5}{c|}{$\times$4}                                                                                                                         \\ \cline{2-11} 
                         & \multicolumn{1}{c}{\textit{EPFL}} & \multicolumn{1}{c}{\textit{HCInew}} & \multicolumn{1}{c}{\textit{HCIold}} & \multicolumn{1}{c}{\textit{INRIA}} & \multicolumn{1}{c|}{\textit{STFgantry}} & \multicolumn{1}{c}{\textit{EPFL}} & \multicolumn{1}{c}{\textit{HCInew}} & \multicolumn{1}{c}{\textit{HCIold}} & \multicolumn{1}{c}{\textit{INRIA}} & \multicolumn{1}{c|}{\textit{STFgantry}} \\ \hline
Bicubic                  & 29.74/0.938              & 31.89/0.936                & 37.69/0.979                & 31.33/0.958               & 31.06/0.950                    & 25.26/0.832              & 27.71/0.852                & 32.58/0.934                & 26.95/0.887               & 26.09/0.845                    \\
VDSR \cite{Kim_2016_CVPR}                    & 32.50/0.960              & 34.37/0.956                & 40.61/0.987                & 34.44/0.974               & 35.54/0.979                    & 27.25/0.878              & 29.31/0.882                & 34.81/0.952                & 29.19/0.920               & 28.51/0.901                    \\
EDSR \cite{Lim_2017_CVPR_Workshops}                    & 33.09/0.963              & 34.83/0.959                & 41.01/0.987                & 34.98/0.976               & 36.30/0.982                    & 27.83/0.885              & 29.59/0.887                & 35.18/0.954                & 29.66/0.926               & 28.70/0.907                    \\
RCAN  \cite{zhang2018rcan}                   & 33.16/0.963              & 35.02/0.960                & 41.13/0.988                & 35.05/0.977               & 36.67/0.983                    & 27.91/0.886              & 29.69/0.889                & 35.36/0.955                & 29.80/0.928               & 29.02/0.913                    \\ \hline
LFBM5D  \cite{alain2018light}                  & 31.15/0.955              & 33.72/0.955                & 39.62/0.985                & 32.85/0.970               & 33.55/0.972                    & 26.61/0.869              & 29.13/0.882                & 34.23/0.951                & 28.49/0.914               & 28.30/0.900                    \\
resLF  \cite{zhang2019residual}                  & 33.62/0.971              & 36.69/0.974                & 43.42/0.993                & 35.40/0.980               & 38.35/0.990                    & 28.26/0.903              & 30.72/0.911                & 36.70/0.968                & 30.34/0.941               & 30.19/0.937                    \\
LFSSR  \cite{yeung2018light}                  & 33.67/0.974              & 36.80/0.975                & 43.81/0.994                & 35.28/0.983               & 37.94/0.990                    & 28.60/0.912              & 30.93/0.914                & 36.91/0.970                & 30.59/0.947               & 30.57/0.943                    \\
LF-ATO  \cite{Jin}                 & 34.27/0.976              & 37.24/0.977                & 44.20/0.994                & 36.17/0.984               & 39.64/0.993                    & 28.51/0.911              & 30.88/0.913                & 37.00/0.970                & 30.71/0.948               & 30.61/0.943                    \\
LF-InterNet \cite{wang2020spatialangular}             & 34.11/0.976              & 37.17/0.976                & 44.57/0.995                & 35.83/0.984               & 38.44/0.991                    & 28.81/0.916              & 30.96/0.916                & 37.15/0.972                & 30.78/0.949               & 30.36/0.941                    \\
LF-DFnet  \cite{wang2020light}               & 34.51/0.976              & 37.42/0.977                & 44.20/0.994                & 36.42/0.984               & 39.43/0.993                    & 28.77/0.916              & 31.23/0.920                & 37.32/0.972                & 30.83/0.950               & 31.15/0.949                    \\
MEG-Net \cite{zhang2021end}                 & 34.31/0.977              & 37.42/0.978                & 44.10/0.994                & 36.10/0.985               & 38.77/0.992                    & 28.75/0.916              & 31.10/0.918                & 37.29/0.972                & 30.67/0.949               & 30.77/0.945                    \\
LF-IINet  \cite{liu2021intra}               & 34.73/0.977              & 37.77/0.979                & 44.85/0.995                & 36.57/0.985               & 39.89/0.994                    & 29.04/0.919              & 31.33/0.921                & 37.62/0.973                & 31.03/0.952               & 31.26/0.950                    \\
DPT   \cite{wang2022detail}                   & 34.49/0.976              & 37.35/0.977                & 44.30/0.994                & 36.41/0.984               & 39.43/0.993                    & 28.94/0.917              & 31.20/0.919                & 37.41/0.972                & 30.96/0.950               & 31.15/0.949                    \\
LFT   \cite{LFT}                   & 34.80/0.978              & 37.84/0.979                & 44.52/0.995                & 36.59/0.986               & 40.51/0.994                    & 29.25/0.921              & 31.46/0.922                & 37.63/0.974                & 31.20/0.952               & 31.86/0.955                    \\
DistgSSR  \cite{Wang_2023}               & 34.81/0.979              & 37.96/0.980                & 44.94/0.995                & 36.59/\textbf{0.986}               & 40.40/0.994                    & 28.99/0.919              & 31.38/0.922                & 37.56/0.973                & 30.99/0.952               & 31.65/0.954                    \\
LFSSR-SAV \cite{cheng2022spatial}               & 34.62/0.977              & 37.42/0.978                & 44.22/0.994                & 36.36/0.985               & 38.69/0.991                    & 29.37/0.922              & 31.45/0.922                & 37.50/0.972                & 31.27/0.953               & 31.36/0.951                    \\
EPIT  \cite{EPIT}                   & 34.83/0.978              & 38.23/0.981                & 45.08/0.995                & 36.67/0.985               & 42.17/0.996                    & 29.34/0.920              & 31.51/0.923                & 37.68/0.974                & 31.37/0.953               & 32.18/0.957                    \\
PILFSSR (Ours)        & \textbf{35.12/0.978}     & \textbf{38.51/0.982}       & \textbf{45.30/0.995}       & \textbf{36.83/}0.984      & \textbf{42.22/0.996}           & \textbf{30.22/0.924}     & \textbf{31.70/0.926}       & \textbf{38.04/0.975}       & \textbf{32.40/0.954}      & \textbf{32.44/0.959}           \\ \hline
\end{tabular} }

\end{table*}

\begin{table*}[t]
\centering
\caption{Quantitative comparisons (PSNR/SSIM) of the proposed approach with the state-of-the-arts for $2\times2\rightarrow7\times7$ angular SR. The best results are highlighted in bold and the second-best results are underlined.}
\label{results:ASR} 
\renewcommand\arraystretch{1.60}
\resizebox{1.0\linewidth}{!}{
\begin{tabular}{|c|cccccccc|}\hline
Datasets & Kalantari \etal\cite{Kalantari} &Wu \etal\cite{Wu} &Yeung \etal\cite{Yeung_2018_ECCV} &Jin \etal\cite{Jin2020LearningLF} &FS-GAF~\cite{Jin_Deep2022} &DistgASR~\cite{Wang_2023} & LF-EASR~\cite{Liu} & PILFASR (Ours)\\
\hline
\emph{30scenes} & 41.40 / 0.982 &33.66 / 0.918&42.77 / 0.986&42.54 / 0.986&42.75 / 0.986 &\underline{43.67} / \underline{0.989} &43.44 / \underline{0.989} &{\bf43.81} / {\bf0.990}\\
\emph{Occlusions} & 37.25 / 0.972 &32.72 / 0.924&38.88 / 0.980&38.53 / 0.979&38.51 / 0.979 &39.46 / 0.983&\underline{39.80} / \underline{0.985} &{\bf40.18} / {\bf0.986}\\
\emph{Reflective} & 38.09 / 0.953 &34.76 / 0.930&38.33 / 0.960&38.46 / 0.959&38.35 / 0.957&39.11 / \underline{0.960} &\underline{39.35} / {\bf0.963} &{\bf39.61} / {\bf0.963} \\
\hline
\emph{HCInew}& 32.85 / 0.909 &26.64 / 0.744 &32.30 / 0.900&34.60 / 0.937&\underline{37.14} / \underline{0.966} &35.96 / 0.959 &35.86 / 0.956 &   {\bf37.17} / {\bf0.971}      \\
\emph{HCIold}   & 38.58 / 0.944 &31.43 / 0.850 &39.69 / 0.941&40.84 / 0.960&41.80 / \underline{0.974} &\underline{42.18} / 0.967  &41.54 / 0.960       &{\bf43.29} / {\bf0.984}  \\
\hline
\end{tabular}}
\end{table*}

\subsection{Network Architecture}\label{subsec:Network architecture}
The overall network architecture is shown in Fig. \ref{network-overview}. To learn the ensemble representation of $C_4^2 $ LF subspaces, we cascade the six subspace-specific feature extractors with 64 channels after a 3$\times$3 spatial convolution. Then, six EPIXformers are subsequently followed to equip the model with geometry-aware decoding capacity.  Finally, we rearrange channel dimension to spatial or angular dimension to have the target size. Specifically, for the SSR task, we use a simple convolution and shuffle net to recover the pixel numbers. For the ASR, we follow~Ref. \cite{Wang_2023,Liu} to achieve angular SR on macro-pixels, 
which recovers the angular resolution by shuffling on the MacPI subspace.

\section{Experiments and Results}

\subsection{Settings}

For spatial SR, we follow previous works \cite{Wang_2023,EPIT,liu2021intra} to use 5 public LF datasets (\textit{i.e.}, \textit{EPFL} \cite{rerabek2016new}, \textit{HCInew} \cite{honauer2017dataset}, \textit{HCIold} \cite{wanner2013datasets}, \textit{INRIA} \cite{le2018light}, \textit{STFGantry} \cite{vaish2008new}) for both training and testing.
We extracted the central $5\times5$ views in each LF image, which were cropped into patches with spatial size of $64\times64$ for $\times 2$ SR and  $128\times 128$ for $\times 4$ SR. 
For each pair, we first convert RGB into YCbCr, then use the Y channel for training or test. We get the LR images resized from the HR ones using bicubic interpolation. The parameter $d_{\textrm{max}}$ of the mask in Eq. (\ref{eq:mask}) is set to 2 pixels for SSR since the maximum disparity of ground-truth LF dataset is about 3 pixels for these datasets and the disparities of the input LF images are further reduced due to the spatial downsampling.

For angular SR, we perform experiments on both real-world and synthetic scenes. We follow previous work to utilize the 100 real-world scenes captured by Lytro Illum from Stanford Lytro Archive~\cite{vaish2008new} and Kalantari \etal~\cite{kalantari2016learning} to train the network.
The real-world test sets are composed of \emph{Occlusion} and \emph{Reflective} splits of Stanford Lytro Archive and \emph{30scenes} from~\cite{kalantari2016learning}.
For synthetic scenes, 20 scenes for HCInew datasets~\cite{honauer2017dataset} are utilized as training data. 
As for the test data, 4 scenes from HCInew~\cite{honauer2017dataset} and 5 scenes from HCIold~\cite{Wanner2013DatasetsAB} are adopted. In this work, we focus on the $2\times2\rightarrow7\times7$ task, since the selected $2\times 2$ SAIs are located at the corner of a $7\times 7$ SAI array, which means that the angular sampling stride is 6 angles,  so the augular upsampling factor $\beta=6$.
To generate the training and testing pairs, we follow previous works~\cite{Kalantari,Yeung_2018_ECCV,Liu,Jin2020LearningLF,Wang_2023} to sample the sparse views from the ground-truth LF images. For ASR, the parameter $d_\textrm{max}$ of the mask in Eq. (\ref{eq:mask}) is set to 6 (18) pixels for real (synthetic) datasets because the maximum disparity of real (synthetic) datasets is about 1 (3) pixels and the disparities in the input LF images are enlarged due to the angular downsampling with a stride of 6 angles.

The data augmentation strategies we use are random horizontally flipping, vertically flipping, $90^o$ rotation.  The networks were trained using $L_1$ Loss, optimized using Adam \cite{kingma2014adam} optimizer ($\beta_1=0.9, \beta_2=0.999$) with a start learning rate $2\times 10^{-4}$ which halfed every 15 epochs. The training was stopped after 80 epochs. The batch size we use is 8/4 for spatial/angular SR.

In the test stage, we reflectively padded the image to deal with the margin region, then cropped the input LF image into patches.
Following previous work~\cite{Wang_2023,EPIT,liu2021intra}, We calculate the peak signal-to-noise ratio (PSNR) and structual similarity (SSIM) metrics on the Y channel. 
In detail, we first calculate the metrics for each SAI of each LF image, then average $A^2$ angles as the PSNR or SSIM for this LF scene (for angular SR, the input SAIs are not used to calculate the metrics), then the PSNR or SSIM of one dataset is the average of all scenes in this test dataset.  See our code for more details.

\subsection{Quantitative Results}
To demonstrate the effectiveness of the proposed method, we compare our spatial SR scheme with 17 other state-of-the-art networks and angular super-resolution scheme with 7 other ones. 
The previous spatial super-resolution works include five single-image SR methods, \textit{i.e.}, bicubic interpolation,  
VDSR \cite{Kim_2016_CVPR}, EDSR \cite{Lim_2017_CVPR_Workshops}, and RCAN \cite{zhang2018rcan}, and 13 LFSSR methods, \textit{i.e.}, LFBM5D \cite{alain2018light}, resLF \cite{zhang2019residual}, LFSSR \cite{yeung2018light}, LF-ATO \cite{Jin}, LF-InterNet \cite{wang2020spatialangular}, LF-DFnet \cite{wang2020light}, MEG-Net \cite{zhang2021end}, LF-IINet \cite{liu2021intra}, DPT \cite{wang2022detail}, LFT \cite{LFT}, DistgSSR \cite{Wang_2023}, LFSSR-SAV \cite{cheng2022spatial}, and EPIT \cite{EPIT}.
The compared angular SR works include Kalantari \etal\cite{Kalantari},  Wu \etal\cite{Wu}, Yeung \etal\cite{Yeung_2018_ECCV}, Jin \etal\cite{Jin2020LearningLF}, FS-GAF~\cite{Jin_Deep2022}, DistgASR~\cite{Wang_2023}, LF-EASR~\cite{Liu}. 
For a fair comparison, all the compared methods except angular super-resolution work \cite{Wu} (the training codes are not publicly available) are trained on the same training datasets as ours.
\subsubsection{Results for LFSSR}
For spatial SR, the results are reported in Table~\ref{SSR-psnr-ssim}.
Compared with the single image SR methods, the performance of LFSSR methods can be highly improved by considering the LF property.
By incorporating the physical prior of LF optical imaging, our method achieved state-of-the-art performance for both $\times 2$ and $\times 4$ spatial SR. 
Compared with 16 previous methods,  we can observe that our method has advantages on real-world \textit{EPFL} and \textit{INRIA} test data, which were captured using a Lytro LF camera. 
Our method outperforms the second-best method EPIT by 0.88 dB on \textit{EPFL} and 1.03 dB on \textit{INRIA} for $\times 4$ SR. That suggests that our method can exploit more information from intrinsic LF prior especially when disparity is very small.
Our method also achieves the best performance on synthetic \textit{HCI} and \textit{STFgantry} test data, in which the data have a relatively larger disparity.
These demonstrate that our method can also handle LF images with large disparity. 
\subsubsection{Results for LFASR}
The quantitative comparison results for angular SR are listed in Table~\ref{results:ASR}. We can observe that our method achieves state-of-art performance on both real-world and synthetic test sets.
Compared with previous methods, for real-world test sets, we can observe that our method outperforms LF-EASR by 0.37 dB on \emph{30scenes} test sets. 
The EPI based method, Wu \etal\cite{Wu} leads to inferior performance compared with depth and non-depth based methods due to the under-utilized spatial information.
The depth-based methods, Kalantari \etal\cite{Kalantari}, Jin \etal\cite{Jin2020LearningLF}, and FS-GAF~\cite{Jin_Deep2022} struggle to handle small-disparity real-world scenes.
As for \emph{Occlusions} and \emph{Reflective} test sets, which contain challenging occluded regions and non-Lambertian surfaces, our method surpasses the second-best method, LF-EASR~\cite{Liu} by 0.38 dB and 0.26 dB, respectively.
For synthetic test sets, our method is the only non-depth based method that outperforms FS-GAF~\cite{Jin_Deep2022} on the HCInew test set. 
Compared with DistgASR~\cite{Wang_2023}, Our method also achieves a gain of 1.11 dB on the HCIold test set.
\subsubsection{Discussion}
The disparity is one of the most important attributes for LF images. For one specific LF data, if the original disparity is at most 4 pixels between every 2 adjacent SAIs, then for $\times 4$ spatial downsampled LF, the disparity is at most $4 / 4=1$ pixel. Similarly, the disparity is at most $4 / 2=2$ pixel for $\times 2$ downsampled one. While for angular SR task, the selected $2\times 2$ SAIs are located at the corner of a $7\times 7$ SAI array, which means that the angular sampling stride is 6 angles and the spatial resolution is not downsampled. 
Therefore, the disparity will be at most $4\times 6 = 24$ pixels. By performing the SR task on $\times 4, \times 2$ SSR and $\times 6$ ASR, we verified that the proposed method could generalize to LF tasks with a wide range of disparity.
\subsection{Visual Results}
In this subsection, we provide visual comparisons for spatial and angular SR tasks. 
\subsubsection{Results for LFSSR}
Fig \ref{SSRx4figure} and Fig \ref{SSRx2figure} present the visual comparison for $\times4$ and $\times2$ spatial SR. In each figure, the zoom-in patches of SAIs, EPIs, and VSIs are highlighted using red, orange, and violet boxes, respectively. 
From Fig \ref{SSRx4figure} (a), 
we can observe that our reconstructed image is sharper and have fewer artifacts than other methods. 
Fig \ref{SSRx4figure} (b) shows an indoor scene, from which we could find that only our method recovered the cusp of the leaves.
Fig \ref{SSRx2figure} shows $\times 2$ SR comparison. 
In Fig \ref{SSRx2figure} (a), from the comparison of a synthetic scene \textit{bicycle}, we can see that the red object reconstructed by our method is clearer, the lines we pointed using arrows are separable while other methods are too blur to distinguish these lines.  
Fig \ref{SSRx2figure} (b) shows the results on a real scene \textit{Palais\_du\_Luxembourg}, the color of pointed position on EPI is yellow while other methods appeared gray, and the VSI indicated that our methods recovered more details than others.
In summary, we can observe that our results are more distinguishable than that of other methods, and more faithful to ground truth.
\subsubsection{Results for LFASR}
The visual comparison results for angular SR are shown in Fig.~\ref{VisualASR}.
In terms of the error map, the results of our method show fewer errors in the foreground objects of scenes \emph{Occlusions\_35\_eslf}. 
From the enlarged patches, we can observe that our method can reconstruct fine-granular details, while the compared methods suffer from obvious artifacts.
Our reconstructed EPIs and VSIs show fewer artifacts than compared methods.
These demonstrate the effectiveness of our method.
\subsection{Ablation Results}
\begin{table}[t]
\caption{Addon to exsiting methods operated on $2\times$ SSR task}
\label{addon}
\renewcommand\arraystretch{1.60}
\resizebox{\linewidth}{!}{
\begin{tabular}{|c|ccccc|c|}
\hline
                         & \multicolumn{1}{c}{\textit{EPFL}} & \multicolumn{1}{c}{\textit{HCInew}} & \multicolumn{1}{c}{\textit{HCIold}} & \multicolumn{1}{c}{\textit{INRIA}} & \multicolumn{1}{c|}{\textit{STFgantry}}  \\ \hline

DistgSSR \cite{Wang_2023}                & 34.81/0.979              & 37.96/0.980                & 44.94/0.995                & 36.59/0.986               & 40.40/0.994 \\
\emph{DistgSSR + VSI}                & 34.89/0.977              & 37.96/0.980                & 44.96/0.995                & 36.75/0.984               & 40.53/0.994  \\
EPIT   \cite{EPIT}                  & 34.83/0.978              & 38.23/0.981                & 45.08/0.995                & 36.67/0.985               & 42.17/0.996  \\

\emph{$C_4^2$ FE + EPIT}              & 34.98/0.979              & 38.50/0.981                & 45.24/0.995                & 36.75/0.986               & 42.00/0.996  \\
Ours               & 35.12/0.978     & 38.51/0.982       & 45.30/0.995       & 36.83\textbf{/}0.984      & 42.22/0.996 \\ \hline
\end{tabular}
}
\end{table}

To investigate the improvement brought by $C_4^2$ feature representations and EPIXformer, we introduced some additional variants.
Firstly, some previous works \cite{Wang_2023, EPIT} can be viewed as our ablation versions. 
Secondly, we separately removed them and retrained the rest part of our network to demonstrate the effectiveness of 2 parts.

\begin{table}[t]
\caption{Ablation experiments operated on $2\times$ SSR task}
\label{SSR ablation}
\renewcommand\arraystretch{1.60}
\resizebox{\linewidth}{!}{
\begin{tabular}{|c|c|ccccc|}
\hline
   \multicolumn{1}{|c|}{$C_4^2$ FE}  &   \multicolumn{1}{c|}{EPIXformer}                    & \multicolumn{1}{c}{\textit{EPFL}} & \multicolumn{1}{c}{\textit{HCInew}} & \multicolumn{1}{c}{\textit{HCIold}} & \multicolumn{1}{c}{\textit{INRIA}} & \multicolumn{1}{c|}{\textit{STFgantry}}  \\ \hline
$\surd$      &   $\times$        & 34.44/0.978              & 37.61/0.978                & 44.62/0.995                & 36.33/0.985               & 39.89/0.993  \\
$\times$       &   $\surd$                     & 34.92/0.977              & 38.24/0.981                & 45.10/0.995                & 36.67/0.985               & 42.21/0.996  \\
  $\surd$      &   $\surd$          & 35.12/0.978     & 38.51/0.982       & 45.30/0.995       & 36.83\textbf{/}0.984      & 42.22/0.996 \\ \hline
\end{tabular}
}
\end{table}

\subsubsection{\texorpdfstring{$C_4^2$ Feature Representation}{C42 Feature Representation}}
To verify the effectiveness of our $C_4^2$ feature representations, we conduct several experiments.
First, to show the superiority of our $C_4^2$ feature extractor over the disentangling block~\cite{Wang_2023}, in which the VSI feature representations are not considered, we introduce a variant by 
adding the VSI convolution of $C_4^2$ FE to DistgSSR, \textit{i.e.} \emph{DistgSSR + VSI}.
As shown in Table~\ref{addon}, the variant \emph{DistgSSR + VSI} outperforms DistgSSR 
by 0.16 dB on \textit{INRIA} dataset.
That conforms to previous hypothesis that VSI contains rich sub-pixel information especially when disparity  is relatively small, such as the real LF images captured using Lytro camera in \textit{EPFL} and \textit{INRIA}.
Second, we add the $C_4^2$ FE into EPIT to form a variant \textit{$C_4^2$ FE + EPIT}, which can also be regarded as removing the ``X" of our proposed structure. As shown in Table~\ref{addon}, the PSNR is 0.26 dB higher on \textit{HCInew} dataset.
Furthermore, we design a variant by removing $C_4^2$ feature extractor, where only EPIXFormers are deployed.
As shown in Table~\ref{SSR ablation}, the PSNR values are decreased by about 0.2 dB.
\subsubsection{EPIXFormer}
As shown in Table~\ref{addon}, the performance of variant \textit{$C_4^2$ FE + EPIT } has demonstrated the influences of EPIXFormer. 
We further removed all EPIXformers from the proposed structure, as shown in Table~\ref{SSR ablation}, the PSNR is dropped by more than 0.5 dB for all test data, especially STFgantry. The reason is that the $C_4^2$ FE in the proposed network had fewer blocks than the variant \textit{DistgSSR + VSI} thus the performance is limited, but it is still competitive to many convolution-based methods, such as LF-DFNet, MEG-Net, and LF-IINet. If we compare variant \textit{EPIXformer} to \textit{DistgSSR + VSI}, we can find that the PSNR values are improved by more than 1.0 dB on \textit{STFgantry} test set which has a large disparity. 
That implies that EPIXformer can help enhance the network's ability to handle large disparities.

\section{Conclusion}

In this work, we propose a method to incorporate the optical imaging prior into LF image SR network design. 
Specifically, the optical imaging process formulation shows two aspects for us: on the whole, LF imaging is 4D imaging with significant 2D constrains, and in detail the LF constrain exhibit an  ``X" shape if we consider every possibility of depth.  Under these descriptions, following the divide-and-conquer philosophy, on the one hand, we complement the VSI subspace to make up the $C_4^2$ subspaces for LF, and correspondingly a $C_4^2$ LF feature extractor to cope with the $C_4^2$ representations 1-by-1. And the proposed VSI contains rich subpixel information especially for LFs with relatively small disparity according to the theoretical analysis as well as experimental results.
On the other hand, we designed EPIXformer to enhance the relevant features that capturing information of the same 3D real world position.
Experimental results show that our method achieves superior performance in both spatial and angular LF super-resolution tasks.
Since our experiments hypothesis that depth information is unknown, so EPIXformer was designed to consider all possibilities. The network can be more delicate if depth information is visible in advance, which may be a solution for future work.

\ifCLASSOPTIONcaptionsoff
  \newpage
\fi

\bibliographystyle{IEEEtran}
\bibliography{ms}

\end{document}